# Language Oriented Modularity: From Theory to Practice

Arik Hadas[a] and David H. Lorenz[a,b]*

a   Open University of Israel
b   Technion—Israel Institute of Technology

**Abstract**   Language-oriented modularity (LOM) is a methodology that complements language-oriented programming (LOP) in providing on-demand language abstraction solutions during software development. It involves the implementation and immediate utilization of domain-specific languages (DSLs) that are also aspect-oriented (DSALs). However, while DSL development is affordable thanks to modern language workbenches, DSAL development lacks similar tool support. Consequently, LOM is often impractical and underutilized.

A challenge for LOM is making the complexity of implementing DSALs comparable to that of DSLs and the effectiveness of programming with DSALs comparable to that of general-purpose aspect languages (GPALs). Today, despite being essentially both domain-specific and aspect-oriented, DSALs seem to be second-class. AspectJ development tools do not work on DSAL code. Language workbenches neither deal with the back-end weaving nor handle the composition of DSALs. DSAL composition frameworks do not provide front-end development tools. DSAL code transformation approaches do not preserve the semantics of DSAL programs in the presence of other aspect languages.

To address this challenge we extend AspectJ with a small set of annotations and interfaces that allows DSAL designers to define a semantic-preserving transformation to AspectJ. We present a transformation approach that enables the use of a standard language workbench to implement DSALs and the use of standard aspect development tools to program with these DSALs. With our approach, DSALs regain their first-class status with respect to both DSLs and aspect languages. This, on the one hand, lowers the cost of developing DSALs to the level of DSLs and, on the other hand, raises the effectiveness of using a DSAL to the level of a GPAL. Consequently, LOM becomes cost-effective compared to the LOP baseline.

As validation, we modified the ajc compiler to support our approach and used it as back-end for two different language workbenches. With Spoofax we implemented Cool to demonstrate that the non-trivial composition of AspectJ and Cool can be accommodated using our approach. With Xtext we applied LOM to crosscutting concerns in two open source projects (oVirt and muCommander), implementing in the process application-specific DSALs, thus providing a sense of the decrease in the cost of developing composable DSALs and the increase in the effectiveness of programming with them.

Crosscutting concerns remain a problem in modern real-world projects (e.g., as observed in oVirt). DSALs are often the right tool for addressing these concerns. Our work makes LOM practical, thus facilitating the use of DSAL solutions in the software development process.

## The Art, Science, and Engineering of Programming

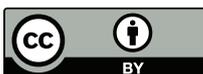

## 1 Introduction

*Language Oriented Modularity (LOM)* [32], taking after *Language Oriented Programming (LOP)* [40, 8], is a programming methodology that involves the development and use of *Domain Specific Aspect Languages (DSALs)* [11] on-demand during the software modularization process. A DSAL is a programming language that is both domain-specific and aspect-oriented. It provides not only domain-specific abstractions and notations like a *Domain Specific Language (DSL)* does, but also a modularization mechanism for the separation of domain-specific crosscutting concerns.

### 1.1 LOM in Theory

In principle, LOM is a special case of LOP, applied to DSALs rather than DSLs.[1] Like LOP, LOM works middle-out. One starts with defining the DSALs and then works outwards, combining high level programming with these DSALs in parallel to their low level implementation. DSALs, like their DSL counterparts, simplify the definition of (crosscutting) domain logic. Moreover, DSALs also simplify the modularization mechanism programmers need to understand (which often makes general-purpose aspect languages (GPALs) more complex to use than the language they extend).

LOM is especially relevant to DSAL-based software development because, in contrast to DSLs, DSALs tend to be tightly coupled with the program for which they were designed. This coupling comes in two forms. First, the weaving specification may rely on the structure of the base code. Second, the crosscutting logic may rely on data retrieved from the base code. This creates a tight coupling between the DSAL code and the representation of the data within the base program. The more coupled the DSAL is with the base code, the less likely it could be reused across applications, thus encouraging ad-hoc application-specific implementation of DSALs [23].

### 1.2 LOM in Practice

In practice, however, LOM is not cost-effective like LOP due to the lack (or incompatibility) of supportive development tools. In terms of cost, *language workbenches* [13] for LOP, such as Xtext [9], MPS [4], Spoofax [26], and Cedalion [35], provide the language designer with a development environment for creating with reasonable effort new DSLs. But these workbenches are not used for LOM because of the inability to express a semantics-preserving transformation of DSALs to existing GPALs [28] (Section 3.2). In terms of effectiveness, language workbenches provide the end-programmer with high quality editing tools for up-to-speed programming with the new DSLs [14]. In contrast, DSALs typically lack similar editing tools and GPAL development tools generally break on DSAL code. Thus the language development and programming experience with DSALs is neither on a par with DSLs nor on a par with GPALs.

---

[1] Unless stated otherwise, we use the term DSL to mean ordinary (non aspect-oriented) domain-specific language rather than its broader meaning which includes DSALs.

### 1.3 Contribution

This work addresses two key challenges that hinder LOM adoption in practice [24, 18, 20]. The first challenge is making DSALs *first-class DSLs* (in the broader sense). By this we refer to the availability of language workbenches for creating DSALs.

The second challenge is making newly created DSALs *first-class aspect languages*. By this we refer to the availability of development tools for programming with DSALs that are normally available when programming with a GPAL. For instance, when one creates a DSAL for a Java [1] application one may expect it to work "out of the box" not only with Java, AspectJ [27], and other DSALs, but also with Eclipse and the *AspectJ Development Tools (AJDT)* [6].

Specifically, we contribute an approach in which DSALs can be implemented like DSLs by transformation to a GPAL (annotated with metadata), without needing to change the compiler (weaver) every time a new DSAL is introduced. This provides an alternative to *aspect composition frameworks* that do require writing compiler code as a part of the DSAL implementation. Making DSAL development more like DSL development allows us to leverage LOP tools for LOM and push down the cost of LOM closer to the LOP baseline. With relatively minor adjustments to GPAL tools, our approach allows us to edit, browse, and compile DSAL code as if it were GPAL code, thus bringing also the effectiveness of LOM closer to LOP [17].

We illustrate our approach concretely with AspectJ. By extending AspectJ with a set of annotations and interfaces, we get a target language for DSALs that brings the cost-effectiveness of the LOM process to the level of LOP. We minimize the additional cost needed to develop DSALs, compared to DSLs, by enabling the use of standard language workbenches also for DSALs. We make programming with DSALs more effective by enabling the use of development tools for AspectJ also for DSALs. Consequently, LOM becomes almost as cost-effective as LOP [19].

### 1.4 Outline

Section 2 motivates by example the need for LOM and the need for defining on-demand DSALs, and explains the difficulty today in applying LOM in practice. Section 3 presents our approach and the key technical idea that enables us to use a GPAL as the target language in ordinary language workbenches. Section 4 describes a concrete implementation of this approach via a set of annotations for AspectJ. The annotations are used for surrendering some obliviousness in return for more control over the visibility of join points and the ordering of advice, and for interfacing with AJDT. In Section 5 we evaluate the improved cost-effectiveness that is achieved by our approach, and apply LOM to crosscutting concerns found in two real-world open source projects.

■ **Listing 1** Core methods of the FileJob class

```
1  public abstract class FileJob implements
       ↪ Runnable {
2    public FileJob(MainFrame mainFrame,
       ↪ FileSet files);
3    public void start();
4    public final void run();
5    protected abstract boolean
       ↪ processFile(AbstractFile file, Object
       ↪ recurseParams);
6    public void interrupt();
7    public void setPaused(boolean paused);
8  }
```

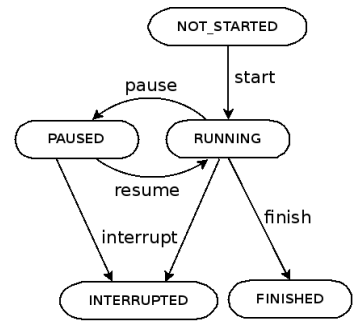

■ **Figure 1** State diagram for a file operation (job)

## 2  Motivation

To illustrate the LOM process and the need for creating DSALs on-demand, let us consider the task of extending a lightweight, cross-platform, open source file manager written in Java, called muCommander,[2] with a mechanism for auditing file operations.

### 2.1  About muCommander

The muCommander file manager supports various operations on files, such as copy, rename and packing, via a dual-pane interface. File operations in muCommander are implemented according to the COMMAND design pattern [15]. Each operation, called *job* in the terminology of muCommander, is encapsulated within a class that extends the abstract class FileJob.

Listing 1 depicts the core methods of the FileJob class. Its constructor receives a set of files on which to operate and the dialog from which the operation was triggered. By calling the *start* method, the logic in the *run* method is executed in a separate thread (hence why FileJob implements the Runnable interface). The abstract method *processFile* is called by the *run* method for each file, and needs to be implemented by subclasses with operation-specific logic. The *interrupt* method stops the execution of a job. Lastly, the *setPaused* method receives a boolean argument *paused,* and accordingly either pauses or resumes the execution of the job.

Figure 1 depicts the permitted state transitions of a job. Initially the state of the job is NOT_STARTED. This state is changed to RUNNING once the job starts executing, and is changed to FINISHED upon completion. If the execution of the job is interrupted (due to an error or by a user request) the state is changed to INTERRUPTED. If the user asks to pause a job, its state is changed to PAUSED. A job that is paused can be either resumed and then its state is changed back to RUNNING or interrupted and then its state is changed to INTERRUPTED.

---

[2] http://www.mucommander.com

■ **Listing 2** Aspect with advice per job

```
1  public privileged aspect Logs {
2      pointcut start() : execution(void start());
3      after(CopyJob job): start() && this(job) { audit(Messages.COPY_STARTED, /* ...skipped... */ ); }
4      after(MkdirJob job): start() && this(job) { /* ...skipped... */ }
5      //... more advice for all combinations of state transitions and job types ...
6  }
```

## 2.2 Implementation Strategies

There are several approaches for implementing a mechanism that audits which jobs were executed in muCommander (including their life-cycle and state transitions).

**Java**    One way is to implement the audit concern in Java. This can be done by extending FileJob with a method for each transition, which concrete job classes could override in order to generate the appropriate messages, and by adding code that persists these messages to the relevant places in FileJob. The drawback of this approach is that it degrades code modularity as the generation of the messages becomes scattered across the job classes and the persistence of these messages becomes tangled among other concerns in FileJob.

**AspectJ**    Another option is AspectJ. Both the generation and the persistence of the messages can be placed within an aspect, separated from the existing code. Listing 2 presents an aspect with an advice that audits the execution of CopyJob (line 3), which copies files from one directory to another. This advice invokes the *audit* method with the message type Messages.COPY_STARTED and the arguments it needs. The *audit* method then retrieves a translated string of the message and places the given values in the right locations to produce a message of the form, e.g., "*start copying 2 files from /home/ to /tmp/ ([/home/a.pdf, /home/b.pdf])*".

The Logs aspect in Listing 2 may seem simple enough, however, there are two issues with using such an aspect in real-world projects. First, it requires all developers in the project to program in AspectJ since new advice needs to be added for every new job. Second, as more jobs are added more pieces of advice are added. That leads to high amount of code duplication and a larger code base that is harder to understand. Note that while putting the common code in an abstract aspect and extending the aspect per job class is an option, it would typically make the process of adding a new job more cumbersome [10].

In order for developers who add jobs not to have to be familiar with AspectJ, one could introduce a single advice per transition instead of having an advice per job class. Inside the body of the advice, the concrete type of the job can be identified and the message type and its values determined accordingly. Developers would only need to modify the body of the advice to change existing audit messages or to add new ones.

However, as more jobs are added to the Logs aspect, the attractiveness of this approach decreases. Listing 3 depicts the Logs aspect with auditing for CopyJob and

■ **Listing 3** Aspect with advice per state transition

```
1  public privileged aspect Logs {
2     pointcut start() : execution(void start());
3     after(FileJob job): start() && this(job) {
4        if (job instanceof CopyJob)
5           audit(Message.COPY_STARTED, /* ... skipped ... */ );
6        else if (job instanceof MkdirJob)
7           if (job.mkfileMode)
8              audit(Message.MKFILE_STARTED, job.files);
9           else
10             audit(Message.MKDIR_STARTED, job.files);
11        //... handle the 'start' transition of other job types ...
12     }
13     //... more advice for other state transitions ...
14  }
```

MkdirJob. The `MkdirJob` class contains a field named *mkfileMode* that determines whether a file or a directory should be created. Since different audit messages are defined for these two cases, this field is checked by the `Logs` aspect (line 7). As more jobs are being audited and as the resolution of the concrete message to be produced gets more complicated, the `Logs` aspect becomes tangled and harder to maintain.

**DSAL Reuse**  One can consider using an off-the-shelf third-party DSAL or reusing a DSAL developed for a similar application, like the one we have implemented for auditing in oVirt,[3] called oVirtAudit [22] (Section 5.2.1). The oVirt platform and the muCommander tool differ in many ways. The former is for virtualization management while the latter is a file manager. The former is a distributed client-server application while the latter is standalone. Lastly, the former is intended for enterprise organizations while the latter is intended for home users. Nevertheless, they both are written in Java, following the COMMAND design pattern, and have operations (called *commands* in the terminology of oVirt) that need to be audited. It is thus tempting to reuse oVirtAudit in muCommander.

However, one quickly discovers that oVirtAudit is not suitable for muCommander. First, the syntax of oVirtAudit does not fit. For example, commands in oVirt cannot be paused and thus oVirtAudit does not provide the syntactic constructs to define a message for an operation that is being paused or resumed. Second, the semantics is different. For example, in oVirtAudit the values placed in the audit messages are taken from method return values while in muCommander they need to be taken from instance variables. Third, the weaving location is different. For example, the advice that produces a message for commands in oVirt that start executing is woven into a method that is not called *start* and is not located within a class named FileJob.

■ **Listing 4** Auditing two file jobs in muCommander using muAudit

```
1   logs for com.mucommander.job.impl.CopyJob:
2       case start log COPY_STARTED with nbFiles baseSourceFolder baseDestFolder files
3       case finish log COPY_FINISHED with nbFiles baseSourceFolder baseDestFolder
4       case interrupt log COPY_INTERRUPTED with baseSourceFolder baseDestFolder
5       case pause log COPY_PAUSED with baseSourceFolder baseDestFolder nbProcessedFiles
6       case resume log COPY_RESUMED with baseSourceFolder baseDestFolder;
7
8   logs for com.mucommander.job.impl.MkdirJob:
9       case start & mkfileMode log MKFile_STARTED with files
10      case start log MKDIR_STARTED with files
11      case finish & mkfileMode log MKFile_FINISHED with files
12      case finish log MKDIR_FINISHED with files
13      case interrupt & mkfileMode log MKFile_INTERRUPTED with files
14      case interrupt log MKDIR_INTERRUPTED with files
15      case pause & mkfileMode log MKFile_PAUSED with files
16      case pause log MKDIR_PAUSED with files
17      case resume & mkfileMode log MKFile_RESUMED with files
18      case resume log MKDIR_RESUMED with files;
```

**New DSAL**  Listing 4 displays the auditing definitions for CopyJob and MkdirJob in muAudit—a simple DSAL we can introduce to allow one to define audit messages in the form of configuration-like *case* statements. The case statements are matched top-down, i.e., the order in which they appear is significant. The first part of each case statement specifies which condition to match based on the job state transition (e.g., *start*) and the values of fields within the job class. The second part defines the message to be produced. This includes the message type (e.g., COPY_STARTED) and the values of fields within the job class. Clearly, muAudit is a declarative and concise way to express the auditing concern. The question is how does one go about creating this DSAL?

### 2.3  Current LOM Solutions

Now that we have established the need to create our own DSAL—muAudit—the supportive development tools are put to the test. We review the support that is currently available for the LOM process with respect to five capabilities (Table 1):

1. *DSAL interoperability*: the ability to define DSALs that can be safely used along with other DSALs;

2. *Development process*: the ability to develop DSALs without requiring compiler (weaver) modifications;

3. *Editing tools*: the ability to produce general editing tools for programming with the DSAL;

4. *Aspect development tools*: the ability to present advice-join-point relationships when browsing the DSAL code; and

5. *Compilation*: the ability to compile DSAL code from the command line.

| Language | Support | CF | LW+GPAL | LW+CF | Practical LOM |
|---|---|---|---|---|---|
| **Domain-specific** | *DSAL interoperability* | ✓ | | ✓ | ✓ |
| | *Development process* | | ✓ | | ✓ |
| **Aspect-oriented** | *Editing tools* | | ✓ | ✓ | ✓ |
| | *Aspect development tools* | | | | ✓ |
| | *Compilation* | | | | ✓ |

■ **Table 1** Comparison of tool support for LOM

**Composition Framework (CF)**   One can implement a DSAL like muAudit with an aspect composition framework such as Awesome [29]. Composition frameworks focus on the implementation of the weaving semantics for DSALs [30]. However, they neither generate general editing tools nor provide the desired browsing and compilation capabilities. Moreover, the need to use compiler development techniques (e.g., bytecode manipulation) results in a highly complex and inefficient development process in terms of LOM.

**Language Workbench + General Purpose Aspect Language (LW+GPAL)**   One may try to implement muAudit using a language workbench. Indeed, the grammar of the DSAL can be defined, and general editing tools can be generated. The natural choice would then be to implement a transformation from that DSAL to a GPAL. This way, we would achieve a DSL-like development process for DSALs. However, a simple transformation of DSALs into a GPAL does not preserve the structure of the code and therefore, in the presence of aspects, does not preserve the meaning of the program. For instance, when the AJAuditor aspect in Listing 5 is used along with the Logs aspect in Listing 7, it will unintentionally expose the executions of the *audit* method that does not exist in the original muAudit code (Listing 4). Such exposure of internal implementation details may result in incorrect behavior of the generated code, e.g., even deadlock [28].[4] Without a transformation, using a language workbench for the development of DSALs is not a viable option. Other than the inability to define the weaving semantics of DSALs, language workbenches also provide neither the desired browsing[5] or debugging capabilities (e.g., lack the original source code location), nor the necessary standalone compilation capabilities.

**Language Workbench + Composition Framework (LW+CF)**   In a sense, language workbenches and composition frameworks are complementary tools. One may consider a simple composition of the two. An efficient way to implement the DSAL would then be to parse it with the language workbench and transform it into the form expected by the composition framework. The expected benefits are: (a) general editing tools

---

[4] Preventing all generated methods (like *audit)* from being advised by modifying all other aspects (like AJAuditor) is impractical, and sometimes impossible in real-world projects.

[5] Browsing capabilities are provided only for the generated GPAL code.

**■ Listing 5** Aspect for auditing method executions

```
1  public aspect AJAuditor {
2      before(): call(* *(..)) && !cflow(within(AJAuditor)) { log("ENTER", thisJoinPoint); }
3      //... skipped ...
4  }
```

are generated by the language workbench; and (b) the DSAL weaving semantics are implemented with a composition framework. Yet, this approach leaves much to be desired [18]. First, it provides browsing capabilities for neither the DSAL code nor the generated code. Second, the need to transform DSAL code before passing it to the composition framework hinders standalone compilation. Third, the development process is highly complex and inefficient due to the need to write compiler code.

### 2.4 Practical LOM

In order to improve the cost-effectiveness of LOM, we seek a solution that supports all the capabilities listed in Table 1. With a first-class development process and DSAL interoperability, the cost of developing DSALs in the context of LOM would become comparable to that of DSLs in LOP. With first-class tools for editing, for aspect development, and for compilation of DSALs, the effectiveness of these DSALs relative to GPALs would become comparable to that of DSLs relative to general-purpose languages in LOP. Ideally, the best practical solution would be to use, when possible, standard language workbenches available for LOP (e.g., Spoofax or Xtext) and standard development tools available for GPALs (e.g., AJDT).

## 3  Approach

In pursuing LOM practicality, LOP is our baseline for comparing cost-effectiveness. In LOP, the cost of DSLs is low thanks to language workbenches (sometimes considered the killer-app for DSLs [13]). Language workbenches enable efficient implementation of DSLs via transformation. Developing a new DSL with a language workbench amounts to writing a transformer (generator) from that DSL to a GPL, and writing a transformer is much easier than writing a compiler or an interpreter. A language workbench also provides tool support for implementing the transformation and for effective editing of DSL code. Once the DSL code is transformed, it is compiled with the GPL's compiler into an executable form, allowing all development tools that are available for the GPL to be used effectively.

To make LOM more practical we facilitate a similar approach to the implementation of DSALs by transforming DSAL code to annotated GPAL code.

### 3.1 Rationale

Implementing DSALs by defining transformations rather than coding their weaving semantics is key to making DSAL creation first-class. First, writing a transformer is much easier than writing a weaver. It eliminates the need to implement a weaver plugin per DSAL, a task that imposes a significant complexity in composition frameworks. Second, when a (semantic-preserving) transformation is possible, the LOM software development process becomes similar to that of LOP and can be completed using existing language workbenches. Indeed, with the Spoofax [26] language workbench one can create for some DSAL an Eclipse plugin that provides editing capabilities (text-highlighting, auto-completion, error-checking, etc.) for writing aspects in that DSAL. Spoofax can also assist in defining a transformation of aspects written in the DSAL into, e.g., AspectJ.

Using a GPAL as the target language is key to also making DSAL development first-class. The transformation of DSALs into a GPAL allows DSAL programmers to leverage, with a one-time adjustment, development tools that exist for the GPAL. These tools work with the transformed code, and the adjustments required for them to provide browsing, navigation, and compilation capabilities for the DSALs are relatively minor.

### 3.2 Pitfalls

Unfortunately, a naive transformation to a GPAL does not work in general. Consider AspectJ as the target language for multiple DSALs. As shown elsewhere [33] translating aspects from different DSALs into aspects in AspectJ and compiling them with the AspectJ compiler (ajc) may yield incorrect behavior (semantic gap). A DSAL for which the transformation to AspectJ is not semantic-preserving becomes a second-class DSL. Meanwhile, trying to fix this by using a different target language and you may lose the prospects of using AJDT for your DSALs (abstraction gap), thus becoming perhaps first-class DSL but second-class aspect language.

**Semantic Gap**   Obliviousness [12] has traditionally been an uncompromising principle in AspectJ. However, in the context of code transformations, complete obliviousness is disadvantageous. In AspectJ the base code cannot refuse advisement (prevent join points from being advised). Consequently, a code transformation that does not preserve the join point "fingerprint" of the original code is not necessarily semantic-preserving in the presence of foreign aspect code (cross-DSAL foreign advising [33]).

Another difficulty is weaving pieces of advice written in different DSALs at the same join point shadow (multi-DSAL co-advising [33]). A conflict occurs when the various pieces of advice are woven in the wrong order. AspectJ provides some control over the ordering of advice by declaring precedence between aspects (via the *declare precedence* statement). However, for programming with multiple DSALs one may need a finer grained ordering mechanism.

**Abstraction Gap**   Development tools for aspect languages heavily rely on the representation of advice-join-point relationships in order to annotate the source code with hints on how aspects are to be woven into the base code. However, code that is generated from DSAL code loses track of the location of advice in the original DSAL code. Consequently, development tools cannot annotate the code, thus hindering effective programming with the DSAL.

In addition, the ability to compile the software from the command line is of a particular interest in real-world projects because of the use of modern tools for continuous integration and continuous delivery. The fact that ajc cannot take DSAL code as input, requires one to modify the compilation process significantly in order to compile the software from the command line.

### 3.3 Bridging the Gap With Metadata

We use metadata to weaken AspectJ in terms of obliviousness, strengthen it in terms of advice ordering, and enhance it in terms of bridging source code locations. The metadata is in the form of Java annotations and an interface for invoking transformations.

**Semantic Gap**    To bridge the semantic gap, a subset of the annotations control the visibility of join points, thus allowing the definition of the transformation to specify where to suppress join point shadows (foreign advising). Another annotation controls the order in which pieces of advice from different DSALs are activated at the same join point shadow (co-advising).

**Abstraction Gap**    To bridge the abstraction gap, metadata can be attached within an annotation, enabling AJDT to provide first-class browsing and navigation capabilities for DSALs. Additionally, DSAL code transformation plugins that implement a special interface are invoked automatically by the compiler in order to provide first-class compilation for DSALs.

## 4    Implementation

We have implemented our approach by modifying the AspectJ compiler (ajc). Our modifications to ajc are both:

- *Optional* - when not in use, the compiler's behavior is unaffected, thus preserving the correctness of the weaving in ajc before the change; and
- *Minimal* - we do the minimal changes necessary to support our extensions, thus we expect the process of reapplying these changes to a newer version of the compiler to be relatively straightforward.

The code changes made to ajc are available at https://github.com/OpenUniversity/ajc and listed in part in Appendix A.

■ **Listing 6**   @Hide annotations added to AspectJ

```
1  @Target(ElementType.FIELD)
2  public @interface HideField {
3      FieldJoinpoint[] joinpoints() default { FieldJoinpoint.SET, FieldJoinpoint.GET };
4  }
5  @Target(ElementType.METHOD)
6  public @interface HideMethod {
7      MethodJoinpoint[] joinpoints() default { MethodJoinpoint.CALL, MethodJoinpoint.EXECUTION,
           ↪ MethodJoinpoint.WITHIN };
8  }
9  @Target(ElementType.TYPE)
10 public @interface HideType {
11    TypeJoinpoint[] joinpoints() default { TypeJoinpoint.PRE_INIT, TypeJoinpoint.INIT,
           ↪ TypeJoinpoint.STATIC_INIT, TypeJoinpoint.WITHIN_INIT,
           ↪ TypeJoinpoint.WITHIN_STATIC_INIT };
12 }
```

## 4.1 Forgoing Complete Obliviousness

Listing 6 defines a set of @Hide annotations for our target language that can be placed on a code element to suppress join point shadows associated with that element:

- @HideField conceals join point shadows associated with a specific field. By default shadows of both *field-set* and *field-get* are hidden, but this can be overridden to be any subset of them.

- @HideMethod conceals join point shadows associated with a specific method. This includes ordinary methods and advice. By default shadows of *method/advice-execution*, *method-call* and all join points that are declared within the method are hidden, but this can be overridden to be any subset of them.

- @HideType conceals join point shadows associated with the initialization of a type. By default shadows of instance *pre-initialization*, instance *initialization*, class *static-initialization*, join points declared within instance *initialization*, and join points declared within *static-initialization* are hidden, but this can be overridden to be any subset of them.

These @Hide annotations are useful when generating AspectJ from DSAL code in order to hide artificial join point shadows in the generated code that do not exist in the original code. Listing 7 illustrates the use of two such annotations (lines 1 and 11).

The support for @Hide annotations is added to ajc by modifying the BcelClassWeaver class, which is responsible for the weaving logic. The decision whether or not to extract a join point is made after inspecting the @Hide annotation, provided that such an annotation exists on the program element with which the join point is associated. For instance, let us consider *method-* and *advice-execution* and join points within a method or advice (Listing 14). When inspecting a LazyMethodGen (which represents an ordinary method or advice), we check for a @HideMethod annotation. If one exists, we retrieve the kind of join points to be hidden. In case of MethodJoinpoint.EXECUTION, we skip matching against the method/advice shadows. In case of MethodJoinpoint.WITHIN,

■ **Listing 7**  Generated aspect from code written in muAudit

```
1  @HideType
2  public privileged aspect Logs {
3      @BridgedSourceLocation(line=1,
           ↪ file="/mucommander/src/main/java/com/mucommander/job/jobs.audit",
           ↪ module="jobs.audit")
4      after(com.mucommander.job.impl.CopyJob job): execution(void start()) && this(job) {
5          if (true) {
6              audit("start_copying_{0}_files_from_{1}_to_{2}_({3})", job.nbFiles, job.baseSourceFolder,
                   ↪ job.baseDestFolder, job.files);
7              return;
8          }
9      }
10     //... skipped ...
11     @HideMethod
12     private void audit(String msg, Object... args) {
13         //... skipped ...
14     }
15 }
```

■ **Listing 8**  @Order annotation added to AspectJ

```
1  public @interface Order {
2      double value();
3  }
```

we skip matching against the join point shadows within the method/advice body. But if the method/advice is not annotated with @HideMethod, we skip nothing.

### 4.2  Fine-Grained Advice Ordering

Listing 8 defines the @Order annotation for ordering pieces of advice *per advice* rather than *per aspect*. The annotation contains a value of type double that represents the precedence of the annotated advice. The lower this value is, the higher the precedence is.

The support for @Order annotations is added to ajc by modifying the *compareTo* method in the BcelAdvice class, which is used to compare pieces of advice (Listing 15). When the advice at hand and the advice with which it is being compared to both have an @Order annotation, the values specified are compared. Otherwise, the comparison defaults to the regular aspect precedence criteria.

### 4.3  Redirect Advice-Join-Point Relations to DSAL Code

In order to leverage the browsing and navigation capabilities of AJDT also for programming with DSALs, the @BridgedSourceLocation annotation is set during the transformation of DSAL code to preserve the source location of advice in the original

■ **Listing 9** @BridgedSourceLocation annotation that was added to AspectJ

```
1  public @interface BridgedSourceLocation {
2      public String file();
3      public int line();
4      public String module();
5  }
```

■ **Listing 10**  Advice in an aspect that was generated from code written in muAudit

```
1  @BridgedSourceLocation(line=9,
        ↪ file="/mucommander/src/main/java/com/mucommander/job/jobs.audit",
        ↪ module="jobs.audit")
2  after(com.mucommander.job.impl.MkdirJob job): execution(void start()) && this(job) {
3      //... skipped ...
4  }
```

DSAL code. The @BridgedSourceLocation annotation (Listing 9) cites a path to a DSAL source code file, a line number, and a module name. When the advice-join-point relationship mapping is returned by the weaver, the source location pointed to by the @BridgedSourceLocation annotation is used instead of the actual location of the advice. This way AJDT markers are shown on the DSAL code and the DSAL code is referred to by markers on advised join points.

Listing 10 shows an example where @BridgedSourceLocation is used. An aspect generated from code written in muAudit specifies that the original source location of *after-execution* advice (line 2) is line 9 in the file */mucommander/src/main/java/com/-mucommander/job/jobs.audit* within the *jobs.audit* module. AJDT uses this information to present the *advises* marker at the right location in the DSAL code and the *advised-by* markers at the locations at which the advice is to be woven (Fig. 2).

The support for the @BridgedSourceLocation annotation is added to ajc by modifying the AsmRelationshipProvider class to retrieve the source location of advice from @BridgedSourceLocation. The main change done to this class is modifying the *getHandle* method to retrieve the @BridgedSourceLocation that annotates the given advice and, if it exists, to return a handle based on the file path and line number it specifies (Listing 16).

### 4.4 Internal Transformation of DSAL Code

In order to compile DSAL code from the command line like ordinary AspectJ code, we require all DSAL-specific transformations to implement an interface called Transformation (Listing 11). This interface declares two methods: the *extension* method returns the file extension of source files that need to be transformed; and the *convert2java* method returns a file containing code generated by the transformation of the DSAL code found within the given file.

An implementation of the Transformation interface enables the compiler to process DSAL code directly. In particular, DSAL code can be compiled from the command

■ **Listing 11** The Transformation interface for DSAL code transformation

```
1  public interface Transformation {
2      String extension();
3      File convert2java(File input) throws Exception;
4  }
```

■ **Listing 12** Implementation of the Transformation interface for muAudit

```
1   public class Main implements Transformation {
2       public static void main(String[] args) { /* ... skipped ... */ }
3       //... skipped ...
4       @Override
5       public File convert2java(File input) throws Exception {
6           Main.main(new String[] { input.getPath() });
7           return new File("src−gen/com/mucommander/job/Logs.aj");
8       }
9       @Override
10      public String extension() { return "audit"; }
11  }
```

line. The compiler invokes the transformation of DSAL code internally. Listing 12 demonstrates the implementation for muAudit. The *extension* method returns *audit* as the extension of muAudit source code files (line 10), and the *convert2java* method calls the transformation of muAudit (line 6) and returns the output file (line 7).

The support for Transformation plugins is added to ajc by modifying the classes AjBuildConfig, ConfigParser, and AjBuildManager. We implemented a pluggable mechanism that uses concrete implementations of Transformation for internal code transformations. First, AjBuildConfig loads all concrete transformations that are specified in a file called *dsals.txt*. Then ConfigParser uses the *extension* method of a loaded transformation in order not to filter out DSAL source files. Just before the compilation process, AjBuildManager transforms the DSAL source files using the loaded transformations (Listing 17).

### 4.5 Discussion

Our approach is based on a transformation of DSALs into a language that extends a GPAL. However, the choice of GPAL is significant. While many of the crosscutting concerns found in real-world projects can be resolved by DSALs whose weaving semantics can be expressed in AspectJ and its join point model, there may be some that are not. Although the join point model of AspectJ could also be extended on-demand, this would likely require much more effort and reduce the cost-effectiveness of the approach. Thus, we intentionally do not consider DSALs that cannot be expressed in AspectJ, but argue that this is a reasonable choice in practice (Section 5).

The design of the @Hide and @Order annotations is based on the classification of multi-DSAL conflicts as foreign- and co- advising [30], and mimics the solution provided by the Awesome composition framework. Nevertheless, our approach is not limited to these particular annotations. The specification of the proposed annotations can be enhanced and even completely replaced with alternative metadata that one can use to resolve multi-DSAL conflicts. Our approach will work as long as the conflict resolution can be specified declaratively and generated during the transformation of DSAL code (with reasonable effort), and as long as it does not impose changes that break the compatibility with the GPAL tools and does not require writing compiler code per DSAL.

It is up to the language designer to determine which shadows to hide using the `@Hide` annotations. Generally, one would want to hide shadows of join points that do not appear in the DSAL code (as these may be considered an internal implementation detail). The `@BridgedSourceLocation` is supposed to be set on every advice in order to present AJDT markers at the right locations for the end-programmer. The more delicate use is that of the `@Order` annotation since the value assigned in the transformation of one DSAL is affected by values assigned in the transformations of other DSALs. Obviously, the `@Order` annotation allows a finer-grained ordering than AspectJ and the use of values of type *double* allows us, theoretically, to introduce right values for newly introduced DSAL without modifying existing DSALs. However, in order to provide a practical way to use the `@Order` annotation, one would probably want to use a tool like SpecTackle [34] to specify the order of all the pieces of advice from all the DSALs that are used in the project, using a user-friendly UI with default resolution. The tool can also help to validate the correctness of specific transformations with business logic tests. However, the implementation of such a tool is out of the scope of this paper.

As we shall see in Section 5, our approach is not coupled with a particular language workbench. One can pick a language workbench that is right for the DSAL task at hand. The only requirement that affects the selection of a language workbench is the need to produce a standalone transformation in order to implement the Transformation interface. However, most mainstream language workbenches provide this capability by default.

A natural question to ask in the context of LOM is what is the complexity of implementing and using a large number of DSALs using our approach. On the one hand, the use of `@Hide` and `@BridgedSourceLocation` annotations and the Transformation interface for each particular DSAL is not affected by the use of other DSALs. On the other hand, for determining the ordering value in the `@Order` annotation for a given advice, one needs to consider pieces of advice defined in other DSALs.

## 5  Evaluation

To evaluate our approach we compare the effort required to implement a complex third-party DSAL with our approach to the effort of implementing the same DSAL in the Awesome composition framework. To assess the impact our approach has on the cost-effectivenss of LOM we present two case studies of implementing and using DSALs in open source projects.

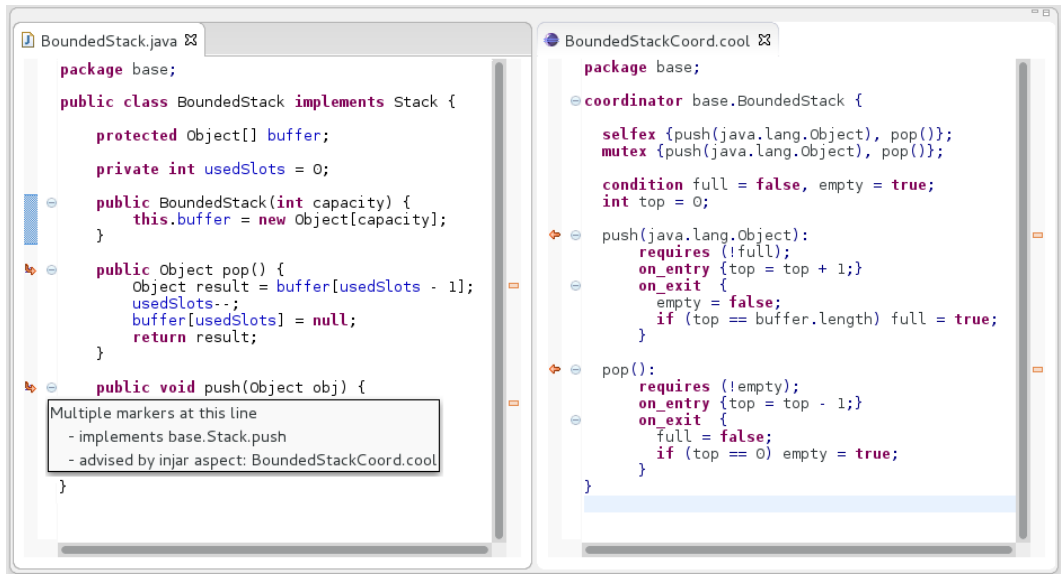

■ **Figure 2** Tool support for programming with Cool in Eclipse

## 5.1 Comparative Evaluation: Implementing Cool

Cool is a DSAL for thread synchronization [31]. It is a non-trivial language to imple-
ment for several reasons. First, a coordinator (aspect in Cool) may include blocks
of code in plain Java. Second, a coordinator has direct access to fields and methods
of the class that is being advised. Third, features of Cool interact with features of
AspectJ [28, 33]. In Cool, join points are not reflected in the syntax, advice comprises
distant terms and expressions, and there is no correct translation to plain AspectJ
(because the implementation-specific operations in the generated aspect might be
ill-advised by other aspects).

We compare two complete implementations of Cool defined in the Spoofax lan-
guage workbench, one following our approach and the other using a composition
framework (CF) approach with Awesome as back-end. In both implementations the
grammar of the language is defined in the *Syntax Definition Formalism (SDF)* [38] (List-
ing 18) and its transformation is implemented in Stratego [39]. In our approach, a
transformation to AspectJ is implemented (with @Hide annotations). In the CF ap-
proach a transformation to Java is implemented, and a new weaver plugin is also
implemented.

To test the interaction of Cool with AspectJ we implemented a coordinator in
Cool (Listing 20) that synchronizes a bounded stack (Listing 19). Figure 2 shows a
screenshot of Eclipse that demonstrates that text-highlighting and AJDT markers are
provided in the process. In addition, we implemented an aspect in AspectJ that audits
method executions and uses the stack (Listing 21). We then ran a multi-threaded
application that reads and writes from and to a bounded stack simultaneously. The
deadlock problem reported elsewhere [28] that occurs when the coordinator is trans-
lated to plain AspectJ was not observed when the @Hide annotations were placed
during the transformation (but reproduced successfully when we removed them).

| Implementation | Grammar | Code Transformation | | Weaver Plugin |
|----------------|---------|---------------------|------|---------------|
| | | EV | Other | |
| Language | SDF | Stratego (AST) | Stratego (AST) | Java |
| CF Approach | 34 | 761 (4168) | 297 (3001) | 1557 |
| Our Approach | 34 | 0 | 382 (3008) | 0 |

■ **Table 2** Number of lines of SDF, Stratego, and Java code in the implementation of Cool

Table 2 compares the implementation effort required when using the two approaches based on #LOC written in three languages. First, the grammar definition in SDF (34 LOC) is the same in both implementations, since the language was defined the same way in Spoofax.

Second, the code transformations are implemented in Stratego. We distinguish between the part of the transformation that handles the resolution of external variables (EV) in Cool and everything else. For each part we compare the #LOC in Stratego and the size of its *Abstract Syntax Tree (AST)* representation in Spoofax. The implementation of the code transformation was much shorter in our approach (382 LOC) than in the alternative approach (761+297=1058 LOC). In our approach, the relatively complex part of the implementation that handles external variables in Cool coordinators was eliminated because it is handled entirely by ajc. On the other hand, the need to generate annotations in our approach slightly increased the size of the other part of the implementation. We argue that the LOC in Stratego is highly affected by code style and therefore the size in LOC of its AST is more representative. The AST of that part using our approach is slightly larger but the difference is insignificant (3008-3001=7).

Third, the implementation of a weaver plugin in Java (1557 LOC) is only required in the CF approach. This emphasizes that in the CF approach one needs to write a relatively large amount of code that is considered to be complicated for most developers since it requires expertise in low level bytecode manipulation tools. In contrast, this knowledge is not needed in our approach since the weaver is not modified. However, the language designer needs to know not only Java but also AspectJ.

## 5.2 Experimental Evaluation

To evaluate our approach in the context of LOM, we present two case studies. In the first case study our approach is used for handling crosscutting concerns found in legacy code of a real-world software. In the second case study our approach is used for handling new requirements on-demand, which is a more typical scenario for LOM.

### 5.2.1 Case Study: oVirt

oVirt is an open source production-ready enterprise application for providing and managing virtual data centers and private cloud solutions. For example, Red Hat

Enterprise Virtualization (RHEV), a commercial competitor to VMware vSphere, is based on oVirt and is deployed in big organizations, such as British Airways.[6]

oVirt-Engine is the control center of the oVirt distributed system that manages the different hosts that run virtual machines. Its core design is based on the COMMAND design pattern [15]. Each operation that is supported by oVirt-Engine is modeled by a command class that inherits from a common root called CommandBase.

We identified three concerns, namely *synchronization*, *auditing*, and *permissions,* that cross-cut many modules in the oVirt-Engine application. The synchronization concern is about preventing conflicting commands from running simultaneously. The auditing concern is about producing informative messages at different stages of command execution. Lastly, the permissions concern is about ensuring that only users with sufficient permissions on entities are able to execute commands that affect them. These concerns are scattered across most of the command classes (as demonstrated in Figs. 3 and 4). In addition, these concerns are tangled within the CommandBase class.

For these crosscutting concerns we implemented in Xtext three DSALs, called oVirtSync, oVirtAudit, and oVirtPermissions, respectively [22]. Their grammars were defined in the language grammar definition format provided by Xtext (Listings 22 to 24). Their transformations into AspectJ with our annotations were implemented in Xtend [3], a language for code transformation provided by Xtext. The transformations of the oVirtSync, oVirtAudit, and oVirtPermissions consisted of 259, 83, and 235 LOC, respectively.[7] With these DSALs, we implemented aspect solutions for three commands named MigrateVmCommand, AddDiskCommand, and ExportVmTemplateCommand (Listings 25 and 26 show aspects written in oVirtSync and oVirtAudit, respectively).

Code scattering was eliminated by encapsulating the code that was spread across the command classes (which, for some commands, exceeded 25% of their LOC) in a single module implemented in the corresponding DSAL. Code tangling was resolved by extracting code that was tangled inside the CommandBase class into the DSAL aspects (296 LOC were untangled, which is more than 12% of the overall LOC). This illustrates that DSALs that are reducible to AspectJ using our approach were effective in separating out the crosscutting concerns we identified in oVirt-Engine. Moreover, the fact that these languages were implemented with Xtext using our approach, unlike our implementation of COOL that was done with Spoofax, validates that our approach is agnostic to the selection of the language workbench. Editing tools and aspect development tools support provided for programming in these languages are demonstrated in Fig. 5.

### 5.2.2 Case Study: muCommander

In this case study the LOM process was applied to the muCommander open source project described in Section 2.

---

[6] http://www.redhat.com/en/about/press-releases/british-airways-chooses-rhev-to-improve-it-systems-to-build-internal-cloud

[7] https://github.com/OpenUniversity/DSALs

■ **Listing 13**   Grammar definition of muAudit in Xtext

```
 1  Model: (commands+=Command)*;
 2  Command:
 3      'logs for' type=[types::JvmDeclaredType|QualifiedName] ':' (cases+=Case)* ';'
 4  ;
 5  Case:
 6      'case' state=State ('&' (fields+=[types::JvmField]))*
 7      'log' msg=[types::JvmEnumerationLiteral] ('with' (vars+=[types::JvmField])+)?
 8  ;
 9  enum State: start | finish | interrupt | pause | resume;
10  QualifiedName: ID ("." ID)*;
```

Using Xtext we implemented a DSAL, named muAudit, for adding a missing auditing feature for job executions in muCommander. We defined the grammar of muAudit in the grammar definition format provided by Xtext (Listing 13). From this grammar definition we generated a plugin for programming with muAudit in Eclipse using Xtext. At this point we were already able to use the IDE plugin and start writing code in muAudit in Eclipse, using general editing capabilities that are typically available when programming with DSLs (Fig. 6). To run muAudit code we implemented a transformation to AspectJ using our approach. This transformation comprised 110 LOC in Xtend. Listing 7 depicts part of the aspect that is generated from the code presented in Listing 4. Note the use of @Hide annotations to hide join point shadows associated with the artificial type *Logs* and its method *audit*.

With the transformation implemented, we were able to use aspect development tools that are typically available for programming with AspectJ, e.g., when writing the code in Listing 4 (Figs. 6 and 7). Finally, we implemented the Transformation interface for muAudit (Listing 12) and added it to the *dsals.txt* file. This enabled us to compile the project not only from within Eclipse but also from the command line, with no changes to the build process.

The LOM development process of muAudit consists of three parts that can be compared to LOP: the grammar definition, the implementation of code transformation, and the implementation of the Transformation interface. The first two parts were done using Xtext, similar to how this is done for DSLs. The third part is also done using Xtext, similar to how it is done when required for DSLs. Even if the third part is typically not needed for DSLs, the additional effort it requires is negligible. Overall, the cost of implementing muAudit is similar to that of a DSL in LOP.

As for the effectiveness of programming with muAudit, not only did we enjoy the benefits of programming with a simplified and more declarative language than AspectJ, but we also enjoyed all the development tools that are usually available for AspectJ. The plugin for Eclipse provided us with general editing tools that are commonly provided by IDEs nowadays. In addition, we were provided with the unique capabilities of aspect development tools by using AJDT and we were able to compile DSAL code the same way we compile AspectJ code. Overall, we were able to program with our DSAL effectively (compared to AspectJ), similar to how one programs with a DSL effectively (compared to Java) in LOP.

### 5.3 Validity and Threats to Validity

The case studies illustrate the effectiveness of our approach in the process of implementing and using DSALs for crosscutting concerns in a real-world project. They demonstrate that with a development effort comparable to that of DSL development (the definition and implementation of the language were done in only a few hours using an existing language workbench) and an effective programming experience comparable to that of a GPAL (existing GPAL tools were used), the cost-effectiveness of the LOM process using our approach is comparable to that of the LOP process.

**Internal Validity**   In LOM the language designer and the DSAL end-programmer are usually different people. In the case studies presented we played both roles. Our familiarity with the DSALs, language workbenches, and the implementation of our extensions to AspectJ could have positively influenced the LOM process. However, to factor out this effect we assess the cost-effectiveness of LOM relative to LOP, comparing the process and tools used.

**External Validity**   One can argue that the DSALs we implemented may not be representative, e.g., that Cool is more complex than most DSALs and that the application-specific DSALs are simpler than most. However, the implementation of Cool is commonly used as a benchmark test case for DSAL frameworks. The fact that the LOM process was cost-effective even for application-specific DSALs is even more impressive than for Cool whose development cost can be amortized across applications.

While being a first-class DSL is a direct consequence of implementing the DSALs with a language workbench, during the case studies we did not test all aspect development tools available for AspectJ, and therefore it is possible that our DSALs are not first-class aspect languages. However, the fact that the browsing capabilities of AJDT (which are typically not available when programming with DSALs) worked in our approach, and that our target language is based on AspectJ, reduces this risk. Yet, it is possible that the @BridgedSourceLocation annotation would need to be enhanced in order to preserve compatibility with future tools.

## 6   Related Work

Various aspect development tools aim at facilitating either the development or the use of DSALs. The Aspect Bench Compiler (abc) [2] is more extensible than the AspectJ compiler (ajc), allowing one to produce extensions to AspectJ and DSALs more easily. However, abc is intended for the development of a particular extension rather than for the composition of extensions. Moreover, abc supports only an old version of AspectJ.

Javassist [5] and similar toolkits that simplify bytecode manipulation can potentially simplify the definition of the DSAL weaving semantics, e.g., when using a composition framework. In contrast, our approach avoids completely the need to program a composition framework extension per-DSAL.

Interpreter-based frameworks like Pluggable AOP [28], JAMI [25], and POPART [7] also avoid low-level implementation of the weaving semantics. However, they achieve simplicity at the expense of performance, since their conflict resolution is based on interpretation. In our approach, the use of DSALs does not imply performance degradation compared to use of GPALs.

The Awesome composition framework [29] generalizes the weaving process of the AspectJ compiler in order to support the definition of DSAL weaving semantics. This approach provides finer-grained constructs for the resolution of foreign advising and co-advising conflicts than the @Hide and @Order annotations we implemented. Our approach, in contrast, provides the ability to define the DSAL weaving semantics without needing low-level bytecode manipulation tools.

The idea of transforming DSALs into AspectJ is found in XAspects [36]. However, a transformation to pure AspectJ does not generally preserve the original meaning of the program. Reflex [37] uses a low-level kernel language to which different aspect languages are transformed. In contrast, our approach fully supports AspectJ and the use of its development tools out-of-the-box when programming with DSALs.

Elsewhere [16, 21] we presented an improved approach to the composition of a language workbench and a composition framework to produce first-class DSALs. In this paper we present an alternative approach, that not only produces first-class aspect languages like the former approach, but also achieves better first-class equality with DSLs by making DSAL development process much more similar to that of a DSL.

SpecTackle [34] is a tool that facilitates the resolution of multi-DSAL co-advising conflicts. It allows one to resolve co-advising conflicts per application, by presenting the conflicts between different aspect languages and allowing programmers to resolve them. This is equivalent to making the values within @Order configurable per application. This topic is left for future research.

## 7    Conclusion

This work addresses the Achilles' heel of LOM practicality, namely that the DSAL development process is far from being cost-effective. On the one hand, DSALs are more costly to develop than DSLs due to the implementation of their weaving semantics. On the other hand, due to the lack of development tools that work on DSALs, the relative effectiveness of programming with a DSAL (relative to a GPAL) is lower than the relative effectiveness of programming with a DSL (relative to a GPL). Consequently, LOM is often avoided or underutilized in practice.

In contrast, LOP is practical thanks to the availability of language workbenches, which provide tools for rapid construction of DSLs as well as for end-programmer productivity in using these DSLs. These tools, however, were generally considered to be inapplicable to DSLs that are aspect-oriented (DSALs). On the one hand, due to the strong coupling between aspects and base code, code transformation may break aspect code or change the meaning of a program as a whole. On the other hand, developing aspect development tools and a weaver for each DSAL is a tedious and complex task.

Our work shows that DSALs can be produced with LOP tools (like language work-benches) and used with GPAL development tools (like AJDT). We present a trans-formation approach that improves the cost-effectiveness of DSAL development and brings it to the level of DSL development. In our approach, code written in DSALs that are reducible to a GPAL can be translated to annotated GPAL code in a semantic-preserving manner and interface with development tools intended for the GPAL. With our approach one can implement DSALs using a standard language workbench, such as Spoofax and Xtext. We present a concrete implementation of the approach with AspectJ as the target language.

In a sense, our work strives to be for DSAL development what the introduction of language workbenches was for DSL development. With our approach, LOM becomes practical for real-world software development process, enabling the on-demand cre-ation and use of DSALs for handling crosscutting concerns, thereby minimizing code scattering and tangling that still prevail in modern software projects.

**Acknowledgment** This research was supported in part by the *Israel Science Founda-tion (ISF)* under grant No. 1440/14.

## A Changes to the AspectJ Compiler

■ **Listing 14** Advice-join-point matching with @Hide annotations in BcelClassWeaver

```
1  private boolean match(LazyMethodGen mg) {
2    //... skipped ...
3    boolean hide = false, hideWithin = false;
4    //... skipped ...
5    AnnotationAJ hideAnn = getHideMethodAnnotation(mg);
6    if (hideAnn != null) {
7      String hideStr = hideAnn.getStringFormOfValue("joinpoints");
8      hide = hideStr == null ||
                ↪ hideStr.contains("Lorg/aspectj/lang/annotation/MethodJoinpoint;EXECUTION");
9      hideWithin = hideStr == null ||
                ↪ hideStr.contains("Lorg/aspectj/lang/annotation/MethodJoinpoint;WITHIN");
10   }
11   //... skipped ...
12   if (canMatchBodyShadows && !hideWithin) {
13     for (InstructionHandle h = mg.getBody().getStart(); h != null; h = h.getNext()) {
14       match(mg, h, enclosingShadow, shadowAccumulator);
15     }
16   }
17   if (canMatch(enclosingShadow.getKind()) && !hide && /** ...skipped ... */ ) {
18     if (match(enclosingShadow, shadowAccumulator)) {
19       enclosingShadow.init();
20     }
21   }
22   //... skipped ...
23 }
```

■ **Listing 15** Advice comparison with @Order annotations in BcelAdvice

```
1  public int compareTo(Object other) {
2    //... skipped ...
3    BcelAdvice o = (BcelAdvice) other;
4
5    try {
6      double diff = getOrder(this) − getOrder(o);
7      if (diff != 0)
8        return diff < 0 ? −1 : 1;
9    } catch (Exception e) {}
10
11   if (kind.getPrecedence() != o.kind.getPrecedence()) {
12     if (kind.getPrecedence() > o.kind.getPrecedence()) {
13       //... skipped ...
14 }
```

**Language Oriented Modularity: From Theory to Practice**

■ **Listing 16** Retrieving source code location with @BridgedSourceLocation annotations in AsmRelationshipProvider

```
1  public static String getHandle(AsmManager asm, Advice advice) {
2    if (null == advice.handle) {
3      AnnotationAJ ann = getSourceLocation(advice);
4      if (ann != null)
5        advice.handle = ann.getStringFormOfValue("file") + ann.getStringFormOfValue("line");
6      else {
7        ISourceLocation sl = advice.getSourceLocation();
8        if (sl != null) {
9          IProgramElement ipe = asm.getHierarchy().findElementForSourceLine(sl);
10         advice.handle = ipe.getHandleIdentifier();
11       }
12     }
13   }
14   return advice.handle;
15 }
```

■ **Listing 17** Invoking DSAL code transformations in AjBuildManager

```
1  public void performCompilation(Collection<File> files) {
2    //... skipped ...
3    for (Iterator<File> fIterator = files.iterator(); fIterator.hasNext();) {
4      File f = fIterator.next();
5      for (Transformation t : AjBuildConfig.transformations)
6        if (f.getName().endsWith(t.extension())) {
7          try {
8            f = t.convert2java(f);
9          }
10         catch (Exception e) {
11           e.printStackTrace();
12         }
13         break;
14       }
15     filenames[idx++] = f.getPath();
16   }
17   //... skipped ...
18 }
```

## B · Grammar Definition for COOL

■ **Listing 18** Grammar definition for COOL in the SDF format

```
1   module languages/cool/Main
2   imports
3       languages/java—15/Main
4       languages/cool/AspectjExtension
5
6   exports
7       sorts
8           ConditionDec CoordinatorBodyDec CoordinatorDec CoordinatorBody MethodAdditionsDec
                   ↪ MethodSignature MutexDec SelfexDec
9
10      context—free syntax
11          CoordinatorDec —> TypeDec
12          CoordinatorDecHead CoordinatorBody —> CoordinatorDec {cons("CoordinatorDec")}
13          "coordinator" TypeName —> CoordinatorDecHead {cons("CoordinatorDecHead")}
14          "{" CoordinatorBodyDec* "}" —> CoordinatorBody {cons("CoordinatorBody")}
15
16          SelfexDec —> CoordinatorBodyDec
17          "selfex" "{" {MethodSignature ","}* "}" ";" —> SelfexDec {cons("Selfex")}
18
19          MutexDec —> CoordinatorBodyDec
20          "mutex" "{" {MethodSignature ","}* "}" ";" —> MutexDec {cons("Mutex")}
21
22          ConditionDec —> CoordinatorBodyDec
23          "condition" {Expr ","}* ";" —> ConditionDec {cons("ConditionDec")}
24
25          MethodAdditionsDec —> CoordinatorBodyDec
26          MethodSignature ":" Requires? OnEntry? OnExit? —> MethodAdditionsDec
                   ↪ {cons("MethodAdditions")}
27          Id "(" {Type ","}* ")" —> MethodSignature {cons("MethodSignature")}
28          "requires" Expr ";" —> Requires {cons("Requires")}
29          "on_entry" Block —> OnEntry {cons("OnEntry")}
30          "on_exit" Block —> OnExit {cons("OnExit")}
31
32          FieldDec —> CoordinatorBodyDec
33
34      lexical syntax
35          "condition" —> Keyword
```

## C The Bounded Stack Example

■ **Listing 19**   Implementation of a non-thread-safe bounded stack

```
1  public class BoundedStack implements Stack {
2     protected Object[] buffer;
3     private int usedSlots = 0;
4     private static BoundedStack instance= null;
5     public BoundedStack(int capacity) { this.buffer = new Object[capacity]; instance=this; }
6     public static BoundedStack getInstance() { return instance; }
7     public Object pop() {
8        Object result = buffer[usedSlots — 1];
9        usedSlots——;
10       buffer[usedSlots] = null;
11       return result;
12    }
13    public void push(Object obj) {
14       buffer[usedSlots++] = obj;
15    }
16 }
```

■ **Listing 20**   Aspect in Cool that synchronizes the bounded stack

```
1  coordinator base.BoundedStack {
2     selfex {push(java.lang.Object), pop()};
3     mutex {push(java.lang.Object), pop()};
4     condition full = false, empty = true;
5     int top = 0;
6     push(java.lang.Object):
7        requires (!full);
8        on_entry {top = top + 1;}
9        on_exit {
10          empty = false;
11          if (top == buffer.length) full = true;
12       }
13    pop():
14       requires (!empty);
15       on_entry {top = top — 1;}
16       on_exit {
17          full = false;
18          if (top == 0) empty = true;
19       }
20 }
```

■ **Listing 21**   Aspect in AspectJ that audits method calls using the bounded stack

```
1  public aspect AJAuditor {
2     pointcut toLog(): call(* *.*(..)) && !cflow(within(org.openu.demo.AJAuditor));
3     before(): toLog() { log(thisJoinPoint); }
4     protected void log(JoinPoint jp) {
5        BoundedStack buf = BoundedStack.getInstance();
6        try { if (buf != null) buf.add(jp); } catch(Exception e) { System.out.println(e.getMessage()); }
7     }
8  }
```

## D  Code Scattering in oVirt

Figures 3 and 4 illustrate scattering of *synchronization*, *auditing*, and *permissions* related code in oVirt, marked in *yellow*, *green*, and *red*, respectively.

```java
public class MigrateVmCommand<T extends MigrateVmParameters> ... {
    private VDS destinationVds;
    private EngineError migrationErrorCode;
    private Integer actualDowntime;
    public MigrateVmCommand(T parameters) { ... }
    public MigrateVmCommand(T migrateVmParameters, CommandContext cmdContext) { ... }
    @Override
    protected LockProperties applyLockProperties(LockProperties lockProperties) { ... }
    public final String getDestinationVdsName() { ... }
    public String getDueToMigrationError() { ... }
    protected VDS getDestinationVds() { ... }
    @Override
    protected void processVmOnDown() { ... }
    protected boolean initVdss() { ... }
    private List<Guid> getDestinationHostList() { ... }
    @Override
    protected void executeVmCommand() { ... }
    private boolean perform() { ... }
    private boolean migrateVm() { ... }
    private MigrateVDSCommandParameters createMigrateVDSCommandParameters() { ... }
    @Override
    public void runningSucceded() { ... }
    protected void getDowntime() { ... }
    private void updateVmAfterMigrationToDifferentCluster() { ... }
    private Boolean getAutoConverge() { ... }
    private Boolean getMigrateCompressed() { ... }
    private int getMaximumMigrationDowntime() { ... }
    private boolean isTunnelMigrationUsed() { ... }
    private String getMigrationNetworkIp() { ... }
    private String getMigrationNetworkAddress(Guid hostId, String migrationNetworkName) {
        ...
    protected boolean migrationInterfaceUp(VdsNetworkInterface nic, List<
        VdsNetworkInterface> nics) { ... }
    @Override
    public AuditLogType getAuditLogTypeValue() { ... }
    private AuditLogType getAuditLogForMigrationStarted() { ... }
    protected AuditLogType getAuditLogForMigrationFailure() { ... }
    protected Guid getDestinationVdsId() { ... }
    protected void setDestinationVdsId(Guid vdsId) { ... }
    @Override
    protected boolean canDoAction() { ... }
    protected void setActionMessageParameters() { ... }
    @Override
    public void rerun() { ... }
    @Override
    protected void reexecuteCommand() { ... }
    protected void determineMigrationFailureForAuditLog() { ... }
    @Override
    protected Guid getCurrentVdsId() { ... }
    public String getDuration() { ... }
    public String getTotalDuration() { ... }
    public String getActualDowntime() { ... }
    @Override
    protected String getLockMessage() { ... }
    private List<Guid> getVdsBlackList() { ... }
    protected List<Guid> getVdsWhiteList() { ... }
    @Override
    public List<PermissionSubject> getPermissionCheckSubjects() { ... }
    @Override
    public void onPoweringUp() { ... }
}
```

■ **Figure 3**  Code scattering in the MigrateVmCommand class

```java
public class AddDiskCommand<T extends AddDiskParameters> ... {
    protected AddDiskCommand(Guid commandId) { ... }
    public AddDiskCommand(T parameters) { ... }
    public AddDiskCommand(T parameters, CommandContext commandContext) { ... }
    @Override
    protected boolean canDoAction() { ... }
    protected boolean checkIfLunDiskCanBeAdded(DiskValidator diskValidator) { ... }
    protected boolean checkIfImageDiskCanBeAdded(VM vm, DiskValidator diskValidator) { ...
        }
    private boolean isShareableDiskOnGlusterDomain() { ... }
    private boolean canAddShareableDisk() { ... }
    private boolean checkExceedingMaxBlockDiskSize() { ... }
    private boolean isStoragePoolMatching(VM vm) { ... }
    protected boolean checkImageConfiguration() { ... }
    private double getRequestDiskSpace() { ... }
    @Override
    protected boolean isVmExist() { ... }
    private DiskImage getDiskImageInfo() { ... }
    private boolean isExceedMaxBlockDiskSize() { ... }
    protected DiskLunMapDao getDiskLunMapDao() { ... }
    protected DiskImageDynamicDao getDiskImageDynamicDao() { ... }
    private Guid getDisksStorageDomainId() { ... }
    @Override
    public Guid getStorageDomainId() { ... }

    @Override
    public List<PermissionSubject> getPermissionCheckSubjects() { ... }

    @Override
    protected void setActionMessageParameters() { ... }
    @Override
    protected void executeVmCommand() { ... }
    private void createDiskBasedOnLun() { ... }
    protected VmDevice addManagedDeviceForDisk(Guid diskId, Boolean isUsingScsiReservation
        ) { ... }
    protected VmDevice addManagedDeviceForDisk(Guid diskId) { ... }
    protected boolean shouldDiskBePlugged() { ... }
    private void createDiskBasedOnImage() { ... }
    private void createDiskBasedOnCinder() { ... }
    private VdcActionParametersBase buildAddCinderDiskParameters() { ... }
    private void setVmSnapshotIdForDisk(AddImageFromScratchParameters parameters) { ... }
    private void addDiskPermissions(Disk disk) { ... }

    @Override
    public AuditLogType getAuditLogTypeValue() { ... }
    private boolean isDiskStorageTypeRequiresExecuteState() { ... }
    private AuditLogType getExecuteAuditLogTypeValue(boolean successful) { ... }
    protected AuditLogType getEndSuccessAuditLogTypeValue(boolean successful) { ... }

    @Override
    protected VdcActionType getChildActionType() { ... }
    protected List<Class<?>> getValidationGroups() { ... }

    @Override
    protected Map<String, Pair<String, String>> getSharedLocks() { ... }
    @Override
    protected Map<String, Pair<String, String>> getExclusiveLocks() { ... }
    @Override
    protected void setLoggingForCommand() { ... }
    private Guid getQuotaId() { ... }
    @Override
    protected void endSuccessfully() { ... }
    private void plugDiskToVmIfNeeded() { ... }
    protected boolean setAndValidateDiskProfiles() { ... }
    @Override
    public List<QuotaConsumptionParameter> getQuotaStorageConsumptionParameters() { ... }
    protected StorageDomainValidator createStorageDomainValidator() { ... }
}
```

■ **Figure 4**   Code scattering in the AddDiskCommand class

**E**    **Grammar Definition for** oVirtAudit, oVirtSync, **and** oVirtPermissions

■ **Listing 22**   Grammar definition for oVirtAudit in Xtext

```
1  Model: (commands+=Command)* ;
2  Command:
3    'logs for' type=[types::JvmDeclaredType|QualifiedName] (overrides?='(overrides)')? ':'
4      (cases+=Case(',' cases+=Case)* (',' 'otherwise' 'log' default=[types::JvmEnumerationLiteral])?)?
5    ';'
6  ;
7  Case:
8    'case' (actionState=[types::JvmEnumerationLiteral] '&')? result=Result ('&' internal?='internal')?
        ↪ ('&' 'state'='(fields+=[types::JvmField]))* ('&' (methods+=[types::JvmOperation]))* 'log'
        ↪ msg=[types::JvmEnumerationLiteral] ;
9  enum Result: success|failure;
10 QualifiedName: ID ("." ID)*;
```

■ **Listing 23**   Grammar definition for oVirtSync in Xtext

```
1  Model: (commands+=Command)* ;
2  Command:
3    'locks for' type=[types::JvmDeclaredType|QualifiedName] '(' scope=Scope (wait?='(& wait')? ')'
        ↪ ':'
4      (exclusiveLocks=Exclusive)? (sharedLocks=Inclusive)? (message=Message)?
5    ';'
6  ;
7  enum Scope: sync|async;
8  Exclusive:
9    {Exclusive} 'exclusively' (override?='(overrides)')? '{' (locks+=Lock(',' locks+=Lock)*)? '}' ;
10 Inclusive:
11   {Inclusive} 'inclusively' (override?='(overrides)')? '{' (locks+=Lock(',' locks+=Lock)*)? '}' ;
12 Lock: 'group: ' group=[types::JvmEnumerationLiteral] 'instance: ' id=[types::JvmOperation]
        ↪ (conditional?='if' condition=[types::JvmOperation])?;
13 Message: 'message: ' type=[types::JvmEnumerationLiteral] (vars+=Var)*;
14 Var: '<' key=STRING ',' value=[types::JvmOperation] '>';
15 QualifiedName: ID ("." ID)*;
```

■ **Listing 24**   Grammar definition for oVirtPermissions in Xtext

```
1  Model: commands+=Command* ;
2  Command:
3    'permissions for' type=[types::JvmDeclaredType|QualifiedName] (extends=Extends)? ':'
4      (permissions+=Permission (',' permissions+=Permission)*)?
5    ';'
6  ;
7  Extends: {Extends} '(extends)' | '(extends if' cond=Condition ')' ;
8  Permission:
9    'object type = ' objectType=[types::JvmEnumerationLiteral] 'object id = '
        ↪ objectId=[types::JvmOperation] 'action group = '
        ↪ actionGroup=[types::JvmEnumerationLiteral] (conditional?='if' conditions+=Condition
        ↪ ('and' conditions+=Condition)*)? ;
10 Condition: (not?='not')? operation=[types::JvmOperation] ;
11 QualifiedName: ID ("." ID)*;
```

**F**    **Aspects for Synchronization and for Auditing in oVirt**

■ **Listing 25**    Synchronizing commands for oVirt expressed in oVirtSync

```
 1  locks for org.ovirt.engine.core.bll.ExportVmTemplateCommand (sync):
 2      exclusively (overrides) { group: REMOTE_TEMPLATE instance: getVmTemplateId }
 3      inclusively (overrides) { group: TEMPLATE instance: getVmTemplateId }
 4      message: ACTION_TYPE_FAILED_TEMPLATE_IS_BEING_EXPORTED <"TemplateName",
         ↪ getVmTemplateName>;

 6  locks for org.ovirt.engine.core.bll.MigrateVmCommand (async):
 7      exclusively (overrides) { group: VM instance: getVmId }
 8      message: ACTION_TYPE_FAILED_VM_IS_BEING_MIGRATED <"VmName", getVmName>;

10  locks for org.ovirt.engine.core.bll.AddDiskCommand (sync):
11      exclusively (overrides) { group: VM_DISK_BOOT instance: getVmId if isBootableDisk }
12      inclusively (overrides) { group: VM instance: getVmId };
```

■ **Listing 26**    Auditing commands for oVirt expressed in oVirtAudit

```
 1  logs for org.ovirt.engine.core.bll.MigrateVmCommand (overrides):
 2      case success & isReturnValueUp log VM_MIGRATION_DONE,
 3      case success & internal log VM_MIGRATION_START_SYSTEM_INITIATED,
 4      case success log VM_MIGRATION_START,
 5      case failure & isHostInPrepareForMaintenance log
         ↪ VM_MIGRATION_FAILED_DURING_MOVE_TO_MAINTENANCE,
 6      case failure log VM_MIGRATION_FAILED;

 8  logs for org.ovirt.engine.core.bll.storage.export.ExportVmTemplateCommand:
 9      case EXECUTE & success log IMPORTEXPORT_STARTING_EXPORT_TEMPLATE,
10      case EXECUTE & failure log IMPORTEXPORT_EXPORT_TEMPLATE_FAILED,
11      case END_SUCCESS & success log IMPORTEXPORT_EXPORT_TEMPLATE,
12      case END_SUCCESS & failure log IMPORTEXPORT_EXPORT_TEMPLATE_FAILED;

14  logs for org.ovirt.engine.core.bll.storage.disk.AddDiskCommand :
15      case EXECUTE & success & internal & isDiskStorageTypeRequiresExecuteState log
         ↪ ADD_DISK_INTERNAL,
16      case EXECUTE & success & isDiskStorageTypeRequiresExecuteState & isVmNameExists log
         ↪ USER_ADD_DISK_TO_VM,
17      case EXECUTE & success & isDiskStorageTypeRequiresExecuteState log USER_ADD_DISK,
18      case EXECUTE & failure & internal & isDiskStorageTypeRequiresExecuteState log
         ↪ ADD_DISK_INTERNAL_FAILURE,
19      case EXECUTE & failure & isDiskStorageTypeRequiresExecuteState & isVmNameExists log
         ↪ USER_FAILED_ADD_DISK_TO_VM,
20      case EXECUTE & failure & isDiskStorageTypeRequiresExecuteState log USER_FAILED_ADD_DISK,
21      case success & isVmNameExists log USER_ADD_DISK_TO_VM_FINISHED_SUCCESS,
22      case success log USER_ADD_DISK_FINISHED_SUCCESS,
23      case failure & isVmNameExists log USER_ADD_DISK_TO_VM_FINISHED_FAILURE,
24      otherwise log USER_ADD_DISK_FINISHED_FAILURE;
```

**G**    **Tool Support for Programming with** oVirtSync

Figure 5 illustrates the general editing tools and aspect development tools that are available while programming with oVirtSync in Eclipse using our approach. An IDE plugin that is generated by Xtext provides one with general editing tools such as text-highlighting, auto-completion (line 17) and syntax-error checking (line 17). In addition, the transformation of the DSAL into AspectJ and the use of the @BridgedSourceLocation annotation enable one to leverage aspect development tools provided by AJDT such as *advises* markers (lines 1 and 8).

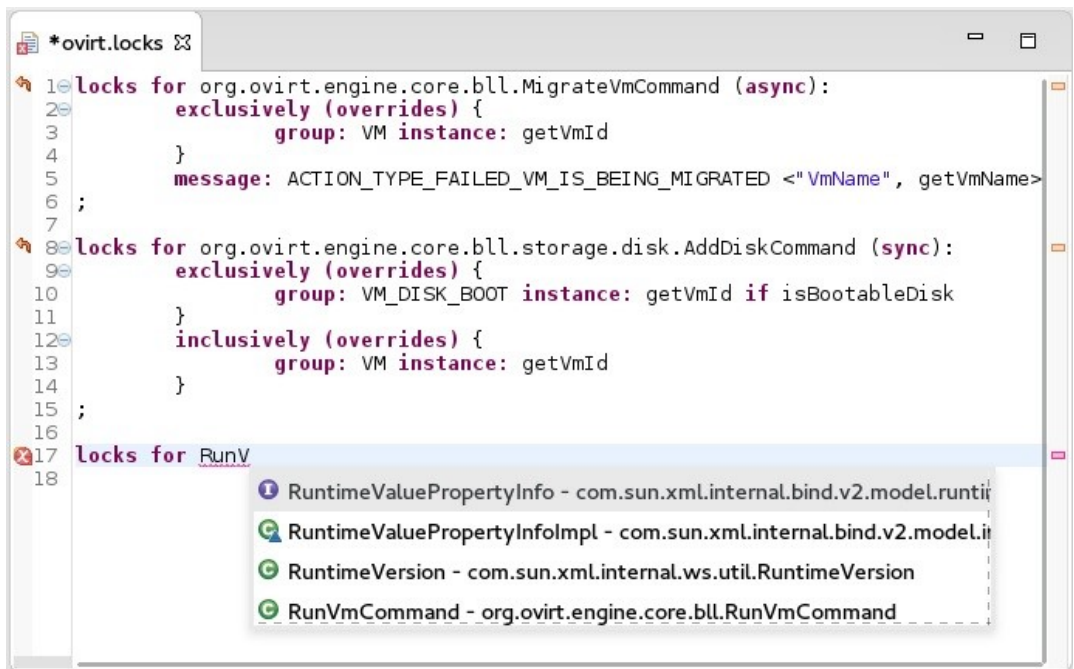

■ **Figure 5**    Aspect in oVirtSync that synchronizes commands in oVirt, edited in Eclipse

**H**   **Tool Support for Programming with** muAudit

```
*jobs.audit ⊠
 1  logs for com.mucommander.job.impl.CopyJob:
 2    case start log COPY_STARTED with nbFiles baseSourceFolder baseDest
 3    case finish log COPY_FINISHED with nbFiles baseSourceFolder baseDe
 4    case interrupt log COPY_INTERRUPTED with baseSourceFolder baseDest
 5    case pause log COPY_PAUSED with baseSourceFolder baseDestFolder nk
 6    case resume log COPY_RESUMED with baseSourceFolder base
 7  ;
 8                                                      baseDestFolder
 9  logs for com.mucommander.job.impl.Mkdir            baseSourceFolder
10    case start & mkfileMode log MKFile_STAR
11    case start log MKDIR_STARTED with files           ;
12    case finish & mkfileMode log MKFile_FIN
13    case finish log MKDIR_FINISHED with fil
14    case interrupt & mkfileMode log MKFile_INTERRUPTED with files
15    case interrupt log MKDIR_INTERRUPTED with files
16    case pause & mkfileMode log MKFile_PAUSED with files
17    case pause log MKDIR_PAUSED with files
18    case resume & mkfileMode log MKFile_RESUMED with files
19    case resume log MKDIR_RESUMED with files
20  ;
```

■ **Figure 6**   Aspect in muAudit that audits jobs in muCommander, editted in Eclipse

```java
    /**
207  * Starts file job in a separate thread.
208  */
209  public void start() {
210
211      // Return if job has already been started
212      if(getState() != FileJobState.NOT_STARTED)
213          return;
214
215      // Pause auto-refresh during file job as it potentially mc
216      // and would potentially cause folder panel to auto-refres
217      getMainFrame().getLeftPanel().getFolderChangeMonitor().set
218      getMainFrame().getRightPanel().getFolderChangeMonitor().se
219
220      setState(FileJobState.RUNNING);
221      startDate = System.currentTimeMillis();
222
223      jobThread = new Thread(this, getClass().getName());
224      jobThread.start();
225  }
226
```

■ **Figure 7**   Code in muCommander being advised by aspect in muAudit

## About the authors

**Arik Hadas** is a software engineer at Red Hat and a graduate student at the Open University of Israel, under the supervision of Prof. David H. Lorenz. His research interests include different aspects of programming and particularly large and distributed software development and modularization techniques. He received his BSc in computer science from the Ben-Gurion University in the Negev, Israel. He is a member of the ACM. Contact him at the Dept. of Mathematics and Computer Science, Open University of Israel, Raanana 4310701, Israel; arik.hadas1@gmail.com. 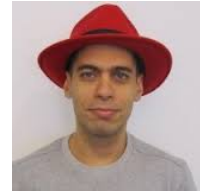

**David H. Lorenz** is an Associate Professor in the Department of Mathematics and Computer Science at the Open University of Israel. He is currently a Visiting Professor at the Faculty of Computer Science, Technion—Israel Institute of Technology. His research interests include language-oriented software engineering, modularity, and programming, particularly involving domain specific languages. Prof. Lorenz received his PhD in computer science from the Technion—Israel Institute of Technology. He is a member of the ACM and a member of the IEEE. Contact him at the Dept. of Mathematics and Computer Science, Open University of Israel, Raanana 4310701, Israel; lorenz@openu.ac.il. 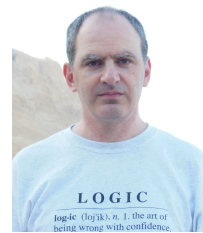